\documentclass[reqno,fleqn]{amsart}
\usepackage{lipsum}
\usepackage[usenames,dvipsnames]{pstricks}
\usepackage{epsfig}
\usepackage{pst-grad} 
\usepackage{pst-plot} 
\usepackage{acronym}
\usepackage{amsmath,amssymb}
\usepackage{graphicx,subfigure}

\newcommand{\R}{\mathbb{R}}
\newcommand{\N}{\mathbb{N}}

\renewcommand{\u}{\mathbf{u}}
\renewcommand{\P}{\mathbb{P}}

\renewcommand{\H}{\mathcal{H}}

\newcommand{\PP}{\mathcal{P}}
\newcommand{\QQ}{\mathcal{Q}}
\newcommand{\Sfrac}[2]{{ \textstyle \frac{#1}{#2}}}

\newcommand{\cf}{cf.~}
\newcommand{\ie}{i.e.~}

\renewcommand{\u}{\bar{u}}

\newcommand{\sech}{\mathrm{sech}}

\newcommand{\half}{{\textstyle{1\over2}}}
\newcommand{\third}{{\textstyle{1\over3}}}

\acrodef{SGN}{Serre-Green-Naghdi}
\acrodef{cB}{`classical' Boussinesq}
\acrodef{DSW}{Dispesive Shock Wave}
\acrodef{DSWs}{Dispesive Shock Waves}
\acrodef{FEM}{Finite Element Method}
\acrodef{IBVP}{initial-boundary value problem}
\acrodef{RK}{Runge-Kutta}
\acrodef{ODEs}{ordinary differential equations}

\begin{document}

\title[Mechanical balance laws for fully nonlinear and weakly dispersive water waves]{Mechanical balance laws for fully nonlinear and weakly dispersive water waves}

\author{Henrik Kalisch}
\address{Department of Mathematics, University of Bergen, Norway}
\email{Henrik.Kalisch@uib.no}

\author{Zahra Khorsand}
\address{Department of Mathematics, University of Bergen, Norway}
\email{Zahra.Khorsand@math.uib.no}

\author{Dimitrios Mitsotakis}
\address{Victoria University of Wellington, School of Mathematics, Statistics and Operations Research, PO Box 600, Wellington 6140, New Zealand}
\email{dimitrios.mitsotakis@vuw.ac.nz}
\urladdr{http://dmitsot.googlepages.com/}



\begin{abstract}
The Serre-Green-Naghdi system is a coupled, fully nonlinear system of dispersive evolution equations which approximates the full water wave problem.
The system is known to describe accurately the wave motion at the surface of an incompressible inviscid fluid in the case when the fluid flow is irrotational and two-dimensional. The system is an extension of the well known shallow-water system to the situation where the waves are long, but not so long that dispersive effects can be neglected. 

In the current work, the focus is on deriving mass, momentum and energy densities and fluxes associated with the Serre-Green-Naghdi system. These quantities arise from imposing balance equations of the same asymptotic order as the evolution equations. In the case of an even bed, the conservation equations are satisfied exactly by the solutions of the Serre-Green-Naghdi system. The case of variable bathymetry is more complicated, with mass and momentum conservation satisfied exactly, and energy conservation satisfied only in a global sense. In all cases, the quantities found here reduce correctly
to the corresponding counterparts in both the Boussinesq and the shallow-water scaling. 

One consequence of the present analysis is that the energy loss appearing in the shallow-water theory of undular bores is fully compensated by the emergence of oscillations behind the bore front. The situation is analyzed numerically by approximating solutions of the Serre-Green-Naghdi equations using a finite-element discretization coupled with an adaptive Runge-Kutta time integration scheme, and it is found that the energy is indeed conserved nearly to machine precision. As a second application, the shoaling of solitary waves on a plane beach is analyzed. It appears that the Serre-Green-Naghdi equations are capable of predicting both the shape of the free surface and the evolution of kinetic and potential energy with good accuracy in the early stages of shoaling.\end{abstract}

\maketitle


\section{Introduction}\label{introduction}
In this paper we study mechanical balance laws for fully nonlinear and dispersive shallow-water waves. 
In particular, the \acf{SGN} system of equations with variable bathymetry is considered. 
This system was originally derived for one-dimensional waves over a horizontal bottom in 1953 
by F. Serre \cite{S1953I, S1953II}. Several years later, 
the same system was rederived by Su and Gardner \cite{Su1969}. 
In 1976, Green and Naghdi \cite{GN1976} derived a two-dimensional fully nonlinear and weakly dispersive 
system for an uneven bottom while in one spatial dimension Seabra-Santos et al. \cite{SRT1987} 
derived the generalization of the Serre system with variable bathymetry. 
Lannes and Bonneton derived several other systems including the \acs{SGN} equations 
using a new formulation of the water wave problem, \cite{LB2009}. For more information 
and generalizations of the \acs{SGN} equations 
we refer to \cite{Lannes2013} and the references therein, 
while we refer to the paper by Barth\'elemy \cite{B2004} for an extensive review. 

The \acf{SGN} system and several variants of it are extensively used in coastal modeling
\cite{wei1995fully,madsen2006boussinesq,Brocchini2013, Lannes2013}.
In the present contribution, the focus is on one aspect of the equations which has not received
much attention so far, namely the derivation and use of associated mechanical balance
equations, and in particular a differential energy balance equation. While it is known
that the equations admit four local conservation equations if the bed is even
\cite{CC2010}, it appears that the connection to mechanical balance laws 
of the original Euler equations has not been firmly established so far.
Here we show that the first three conservation laws of the \acf{SGN} equations
arise as approximations of mechanical balance laws in the context of the Euler equations,
both in the case of even beds, and in the case of nontrivial bathymetry.
The fourth conservation law has been shown to arise from a kinematic identity
similar to Kelvin's circulation theorem \cite{GKK2015}.

Let us first review some modeling issues regarding the \acf{SGN} system.
Suppose $a$ denotes a typical amplitude, and $l$ a typical wavelength
of a wavefield under study.
Suppose also that $b_0$ represents the average water depth.
In order to be a valid description of such a situation, the \acs{SGN} equations 
require the shallow water condition, $\beta \doteq b_0^2/l^2 \ll 1$.
In contrast, the range of validity of the weakly nonlinear 
and weakly dispersive Boussinesq equations is limited
to waves with small amplitude and large wavelength, 
i.e. $\alpha \doteq a/b_0 \ll 1$ and $\beta \ll 1$.
In this scaling regime, one also finds the weakly nonlinear, fully dispersive Whitham equation
\cite{Lannes2013,Whitham67PRSA,EK09}.

The \acs{SGN} equations can be derived by depth-averaging the Euler equations 
and truncating the resulting set of equations at $\mathcal{O}(\beta^{2})$ without making any assumptions 
on the order of $\alpha$, other than $\alpha \le \mathcal{O}(1)$.

In their dimensionless and scaled form the \acs{SGN} equations can be written as
\begin{subequations}\label{eq:serre1}
\begin{align}
& \eta_{t}+[h\bar{u}]_{x}=0\ ,\label{eq:serre1a} \\
& \bar{u}_{t} + \bar{u}\bar{u}_{x} + g \eta_{x}+
\frac{1}{h} \left[ h^{2} \big( \Sfrac{1}{3} \PP + \Sfrac{1}{2} \QQ \big) \right]_x
 + b_x \big( \Sfrac{1}{2} \PP +  \QQ \big) =0 \ ,
\label{eq:serre1b}
\end{align}
\end{subequations}
with
$ \PP =  h \left[ \bar{u}_{x}^{2} - \bar{u}_{xt} - \bar{u} \bar{u}_{xx} \right]$ 
and
$ \QQ = b_x (\bar{u}_t +  \bar{u}\bar{u}_x) + b_{xx} \bar{u}^2$, $x\in\R$, $t>0$, 
along with the initial conditions
$h(x,0)=h_0(x)$, $\u(x,0)=\u_0(x)$. Here, $\eta = \eta(x,t)$ is the free surface displacement, while 
\begin{equation}\label{E2}
h\doteq \eta-b\ ,
\end{equation}
denotes the total fluid depth. The unknown $\u = \u(x,t)$ is the depth-averaged
horizontal velocity, and $\eta_0$, $\u_0$ are given real 
functions, such that $\eta_0-b> 0$ for all $x\in
\R$. In these variables, the location of the horizontal bottom is
given by $z = b$ (cf. Figure 1). For a review of the derivation and the basic
properties of this system we also refer to \cite{B2004,CBB1}.

In the case of small-amplitude waves, \ie if $\beta \sim \alpha$, 
the \acs{SGN} equations reduce to Peregrine's system \cite{P1967}. 
On the other hand, in the case of very long waves, \ie $\beta \to 0$, 
the dispersive terms disappear,
and the system reduces to the nondispersive shallow water equations.

The \acs{SGN} system for waves over a flat bottom possesses solitary and cnoidal wave solutions
given in closed form.
For example, the solitary wave with speed $c_s$ can be written as 
\begin{subequations}\label{eq:solwave}
\begin{align}
  & h_s(\xi) \doteq h_s(x,t)=a_0+a_1{\sech}^2(K_s\, \xi),  \label{eq:solwavea} \\
  & u_s(\xi) \doteq u_s(x,t)=c_s\left(1-\frac{a_0}{h_s(\xi)}\right), \label{eq:solwaveb}
\end{align}
\end{subequations}
where $\xi = x-c_s t$, $K_s = \sqrt{3 a_1/4  a_0^2 c_s^2}$, $c_s = \sqrt{a_0 + a_1}$, 
and $a_0 > 0$ and $a_1 > 0$. For more information about the solitary and cnoidal waves and their dynamical properties we refer to \cite{B2004,CC2010,Li2001,Li2002,MDC2014,Khorsand2014}.

It is important to note that the \acs{SGN} system has a Hamiltonian structure,
even in the case of two-dimensional waves over an uneven bed
\cf \cite{Li2002, CHL1996, Johnson2002, Israwi2011}. Specifically, any solution $(h,\u)$ 
of (\ref{eq:serre1}) conserves the Hamiltonian functional
%
\begin{equation}
\label{eq:ham}
 \H(t) = \frac{1}{2} \int_{-\infty}^{\infty} g\eta^2 + h \u^2 + h\left[h_x b_x + \half h b_{xx} + b_x^2\right] \u^2-
\third\left[h^3 \u_{x}\right]_x \u~ dx\ ,
\end{equation}
in the sense that $d \mathcal{H}(t)/dt=0$. 
Note however that (\ref{eq:serre1a}), (\ref{eq:serre1b}) are recovered only
if a non-canonical symplectic structure matrix is used.
While in many simplified models equations, the Hamiltonian functions does not represent the
mechanical energy of the wave problem \cite{Kalisch2015}, 
in the case of \acs{SGN}, the Hamiltonian
does represent the approximate total energy of the wave system.
Thus the Hamiltonian can be written in the form
\begin{equation*}
\H(t) = \int_{-\infty}^{\infty} E(x,t) \, dx\ ,
\end{equation*}
where the integrand 
\begin{equation*}
E =\frac{1}{2}\left( g \eta^2 +  h\u^2  + h\left[h_x b_x+\half h b_{xx}+b_x^2\right]\u^2
     - \third \left[h^3 \u_{x}\right]_x \u \right)
\end{equation*}
is the depth-integrated energy density.
In the present paper, we also identify a local depth-integrated 
energy flux $q_E$, such that 
an equation of the form
\begin{equation}\label{eq:consenb}
\frac{\partial E}{\partial t}
 + \frac{\partial{q_{E}}}{\partial x} = 0\ , 
\end{equation}
is satisfied approximately.
The procedure of finding the quantities $E$ and $q_E$ follows a similar outline
as the derivations in \cite{AlKa2012JNLS} for a class of Boussinesq systems
and \cite{AK2014} for the KdV equation. 
Expressions for energy functionals associated to Boussinesq 
systems have also been developed in \cite{FKD2014}.

The analytical results are put to use in the study of undular bores.  
It is well known that the shallow-water theory for bores predicts
an energy loss  \cite{Rayleigh1914}.
In an undular bore, the energy is thought to be disseminated through an increasing
number of oscillations behind the bore, and the traditional point of view is that 
dissipation must also have an effect here \cite{BenjaminLighthill1954,Sturtevant1965}. 
However, recent studies \cite{AK2010} have shown that if dispersion is
included into the model equations, then the energy loss experienced 
by an undular bore can be accounted for without making appeal to dissipative mechanisms.

Indeed, it was argued in \cite{AK2010, AK2012} that the energy loss in an undular bore
could be explained wholly within the realm of conservative dynamics by 
investigating a higher-order dispersive system, and monitoring the associated
energy functional. However, there was a technical problem in the analysis
in these works, as the energy functional was not the same as the one required by 
the more in-depth analysis in \cite{AlKa2012JNLS}. On the other hand, the energy functional
found in \cite{AlKa2012JNLS} did not reduce to the shallow-water theory in the correct way.
In the current contribution, it is our purpose to remedy this situation by 
using the \acs{SGN} system which reduces in the correct way to the shallow-water
equations, and also features exact energy conservation in the case of a flat bed.

The numerical method to be used is a standard Galerkin / \acf{FEM} for the \acs{SGN} equations with 
reflective boundary conditions extending the numerical method presented in \cite{MID2013}. 
For the sake of completeness we mention that there are several numerical methods 
applied to boundary value problems of the \acs{SGN} equations. 
For example finite volume \cite{LeMetayer-10,Dutykh2013, Bonneton2011}, 
finite differences \cite{CBB1,CBB2, EG2006},
spectral \cite{Dutykh2013} and Galerkin methods \cite{MID2013,DSM2015}. 
 
The paper is organized as follows: A review of the derivation of the \acs{SGN}  
equations based on \cite{B2004,CBB1} is presented in Section \ref{sec:derivation}. The derivation 
of the mass, momentum and energy balance laws in the asymptotic order of the \acs{SGN}  
equations is presented in Section \ref{sec:laws}.  Applications to undular bores and 
solitary waves are discussed in Section \ref{sec:applications}. 
The numerical method to be used used in this paper is presented briefly in \ref{app:A}.

\section{The \acs{SGN} equations over a variable bed}\label{sec:derivation}
Before introducing the balance laws for the \acs{SGN} equations, 
we briefly review the derivation of the \acs{SGN} equations from the Euler 
equations following the work \cite{B2004}, but in the case of a general bathymetry.
This well known derivation is included here to set the stage for the development
of the approximate mechanical balance laws in the next section.
We consider an inviscid and incompressible fluid,
and assume that the fluid flow is irrotational and two-dimensional.
Let $a_{0}$ be a typical amplitude, $l$ a typical wavelength and $b_{0}$ a typical water depth. 
We perform the change of variables $\tilde{x}=x/l$, $\tilde{z}=z/b_{0}$, $\tilde{t}=c_0t/l$, which yields non-dimensional independent
variables identified by tildes, where $x$ represents the horizontal and $z$ the vertical coordinate.
The limiting long-wave speed is defined by $c_{0}=\sqrt{gb_{0}}$,
and $g$ denotes the acceleration due to gravity. 
The non-dimensional velocity components are defined by
$\tilde{u}=u/\alpha c_{0}$, $\tilde{v}=v/\sqrt{\beta}\alpha c_{0}$,
where $\alpha=a_{0}/b_{0}$ and $\beta=b_{0}^{2}/l^{2}$. 
Finally, the free surface deflection, bottom topography and pressure 
are non-dimensionalized by taking $\tilde{\eta}=\eta/a_{0}$,  $\tilde{b}=b/b_{0}$, and $\tilde{p}=p/\rho g b_{0}$.

\begin{figure}[ht]
  \centering
  \includegraphics[width=4in]{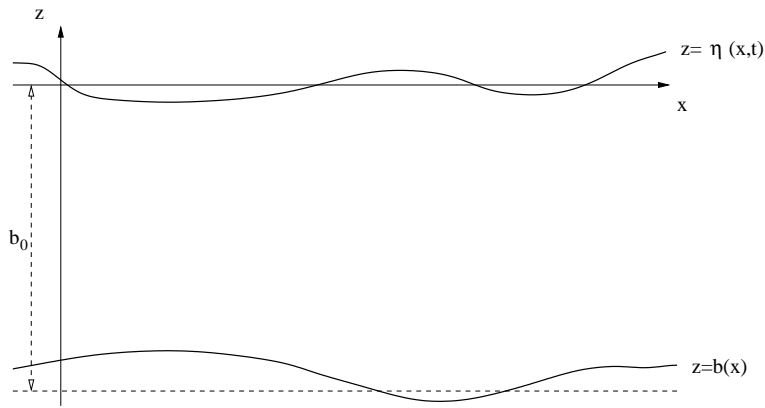}
  \caption{The geometry of the problem}
  \label{fig:figure0}
\end{figure}

In non-dimensional variables, the free-surface problem is written as follows \cite{Whitham1999}:
The momentum equations are
\begin{subequations}\label{eq:euler}
\begin{align}
 &  \alpha \tilde{u}_{\tilde{t}}+\alpha^{2}(\tilde{u}^{2})_{\tilde{x}}+\alpha^{2}(\tilde{u}\tilde{v})_{\tilde{z}} 
    = - \tilde{p}_{\tilde{x}}\ ,  \label{eq:momentum1}\\
 & \alpha \beta \tilde{v}_{\tilde{t}}+\alpha^{2} \beta \tilde{u} \tilde{v}_{\tilde{x}}+\alpha^{2} \beta \tilde{v}\tilde{v}_{\tilde{z}}
    = - \tilde{p}_{\tilde{z}}-1\ .  \label{eq:momentum2}
\end{align}
\end{subequations}
The equation of continuity and the irrotationality are expressed by
\begin{subequations}
\begin{align}
 & \tilde{u}_{\tilde{x}}+\tilde{v}_{\tilde{z}}=0\ ,  \label{eq:continuity1} \\
 &  \tilde{u}_{\tilde{z}}-\beta \tilde{v}_{\tilde{x}}=0\ . \label{eq:irrotationality}
\end{align}
\end{subequations}
The boundary conditions at the free surface and at the bottom are given by
\begin{subequations}
\begin{align}
\tilde{v} &=  \tilde{\eta}_{\tilde{t}} + \alpha \tilde{u}\tilde{\eta}_{\tilde{x}},  
& \mbox{  at  } &\tilde{z}= \alpha \tilde{\eta}(\tilde{x})\ , \qquad\qquad\qquad\qquad\qquad \label{bc1} \\
\tilde{p} &= 0,   & \mbox{  at  } &\tilde{z}=\alpha \tilde{\eta}(\tilde{x})\ , \label{bc2} \\
\tilde{v} &=  \tilde{b}_{\tilde{x}} \tilde{u},   & \mbox{  at  } &\tilde{z}=\tilde{b}(\tilde{x})\ . \label{bc3}
\end{align}
\end{subequations}

The first equation in the system \eqref{eq:serre1} is obtained by integrating
the equation of continuity over the total depth. The result is written in terms
of the depth-averaged horizontal velocity 
\begin{equation}\label{eq:depav}
\bar{\tilde{u}}=\frac{1}{\tilde{h}}\int_{\tilde{b}}^{\alpha\tilde{\eta}}\tilde{u}~dz\ ,
\end{equation}
in the form
\begin{equation}\label{eq:continuity2}
 \tilde{\eta}_{\tilde{t}}+[\tilde{h}\bar{\tilde{u}}]_{\tilde{x}}=0\ . 
\end{equation}
Using the boundary conditions (\ref{bc1})-(\ref{bc3}), 
the continuity equation (\ref{eq:continuity2}) 
and the depth-averaged momentum equation (\ref{eq:momentum1}) yields
\begin{equation}\label{eq:momentum3}
\alpha \tilde{h} \bar{\tilde{u}}_{\tilde{t}}+\alpha^{2} \tilde{h} \bar{\tilde{u}}\bar{\tilde{u}}_{\tilde{x}}
 + \alpha^{2}\frac{\partial}{\partial \tilde{x}}
  \int_{\tilde{b}}^{\alpha \tilde{\eta}}
 \left(\tilde{u}^{2}-(\bar{\tilde{u}})^{2}\right)\,d\tilde{z}
= -\int_{\tilde{b}}^{\alpha \tilde{\eta}} \tilde{p}_{\tilde{x}}\,d\tilde{z}\ . 
\end{equation}
Applying the Leibniz rule to the right-hand side of equation (\ref{eq:momentum3}) yields
\begin{align}
 \int_{\tilde{b}}^{\alpha \tilde{\eta}} \tilde{p}_{\tilde{x}}\,d\tilde{z}
&= \frac{\partial}{\partial \tilde{x}} \big( \tilde{h} \bar{\tilde{p}} \big) 
                                                       - \alpha \eta_{\tilde{x}} \tilde{p}|_{\tilde{z}= \alpha \tilde{\eta}} 
                                                       +  \tilde{b}_{\tilde{x}} \tilde{p}|_{\tilde{z}=\tilde{b}} \nonumber \\
&= \frac{\partial}{\partial \tilde{x}} \big( \tilde{h} \bar{\tilde{p}} \big)  
                     +  \tilde{b}_{\tilde{x}} \tilde{p}|_{\tilde{z} = \tilde{b}} \ .  \label{eq:momentum4}
\end{align}

The momentum equation (\ref{eq:momentum2}) is rewritten as 
\begin{equation}\label{eq:pressure1}
\alpha \beta \Gamma(\tilde{x},\tilde{z},\tilde{t}) = -1 - \tilde{p}_{\tilde{z}} \ ,
\end{equation}
where
\begin{equation}\label{eq:gamma}
 \Gamma(\tilde{x},\tilde{z},\tilde{t})
= \tilde{v}_{\tilde{t}}+\alpha \tilde{u}\tilde{v}_{\tilde{x}}+\alpha\tilde{v}\tilde{v}_{\tilde{z}}\ .
\end{equation}
Integrating equation (\ref{eq:pressure1}) from $\tilde{z}$ to $\alpha \tilde{\eta}$ yields
\begin{equation}
\tilde{p}(\tilde{x},\tilde{z},\tilde{t})=(\alpha \tilde{\eta} - \tilde{z})
+\alpha \beta \int^{\alpha\tilde{\eta}}_{\tilde{z}} \Gamma(\tilde{x},\zeta,\tilde{t})\,d\zeta \ , \label{eq:pressure2}
\end{equation}
and taking the mean value gives
\begin{equation}\label{eq:pres}
\tilde{h}\bar{\tilde{p}} = \frac{1}{2}\tilde{h}^{2}
+ \alpha \beta \int_{\tilde{b}}^{\alpha \tilde{\eta}} \int_{\tilde{z}}^{\alpha \tilde{\eta}} 
\Gamma(\tilde{x},\zeta,\tilde{t})\, d\zeta\  d\tilde{z} \ .
\end{equation}
Therefore, equation (\ref{eq:momentum3}) can be written as
\begin{multline*}
\bar{\tilde{u}}_{\tilde{t}} + \alpha \bar{\tilde{u}}\bar{\tilde{u}}_{\tilde{x}} + \tilde{\eta}_{\tilde{x}}
+ \frac{\beta}{\tilde{h}} \frac{\partial}{\partial \tilde{x}}\int_{\tilde{b}}^{\alpha \tilde{\eta}} 
 (\tilde{z} - \tilde{b}) \Gamma(\tilde{x},\tilde{z},\tilde{t})\,d \tilde{z}
\\
+ \frac{\beta}{\tilde{h}} 
   \tilde{b}_{\tilde{x}} \int_{\tilde{b}}^{\alpha \tilde{\eta}} \Gamma(\tilde{x},\tilde{z},\tilde{t})\,d \tilde{z}
=
\frac{-\alpha}{\tilde{h}}\frac{\partial}{\partial \tilde{x}} \int_{\tilde{b}}^{\alpha \tilde{\eta}}
 \left(\tilde{u}^{2}-(\bar{\tilde{u}})^{2}\right)\, d\tilde{z}\ .
\end{multline*}
The non-dimensional velocity componens are given (\cf \cite{CBB1}) to first order by
\begin{equation}\label{eq:lowhvel1}
\tilde{u}(\tilde{x},\tilde{z},\tilde{t})= \bar{\tilde{u}}(\tilde{x},\tilde{t}) +\mathcal{O}(\beta)\ ,
\end{equation}
and
\begin{equation}\label{eq:lowvvel1}
\tilde{v}(\tilde{x},\tilde{z},\tilde{t}) 
= - \big( \tilde{z} - \tilde{b}(\tilde{x}) \big) \frac{\partial\bar{\tilde{u}}}{\partial \tilde{x}}
  + \bar{\tilde{u}} \frac{\partial \tilde{b}}{\partial \tilde{x}}
+\mathcal{O}(\beta)\ .
\end{equation}

As it was shown in \cite{CBB1}, we can expand the velocity components 
using Taylor series in the vertical coordinate around the bottom. 
Denoting by $\tilde{u}^{b}$ and $\tilde{v}^{b}$, respectively, the horizontal and vertical velocities at the bottom, 
the bottom kinematic condition (\ref{bc3}) imposes that $\tilde{v}^{b}=\tilde{b}_{\tilde{x}}\tilde{u}^{b}$. 
In order to determine which terms should be kept to obtain an approximation for the velocity field, 
the incompressibility condition (\ref{eq:continuity1}) must hold to the 
same order in $\beta$ as the evolution equations. If the non-dimensional velocity components are given by
\begin{equation}
\tilde{u}(\tilde{x},\tilde{z},\tilde{t})= \tilde{u}^{b}(\tilde{x},\tilde{t})+ \beta (\tilde{z}-\tilde{b})\left(\tilde{b}_{\tilde{x}}\tilde{u}_{\tilde{x}}^{b}+
(\tilde{b}_{\tilde{x}}\tilde{u}^{b})_{\tilde{x}}\right)\\
- \frac{\beta}{2}(\tilde{z}-\tilde{b})^{2}\tilde{u}_{\tilde{x}\tilde{x}}^{b}+\mathcal{O}(\beta^{2})\ , \label{eq:hvelocity1}
\end{equation}
\begin{multline*}
\tilde{v}(\tilde{x},\tilde{z},\tilde{t})= \tilde{b}_{\tilde{x}}\tilde{u}^{b} + (\tilde{z}-\tilde{b})\left(-\tilde{u}_{\tilde{x}}^{b}+
\beta (\tilde{b}_{\tilde{x}}(\tilde{u}^{b}\tilde{b}_{\tilde{x}})_{\tilde{x}}+ \tilde{u}_{\tilde{x}}^{b}\tilde{b}_{\tilde{x}}^{2})\right)\\-
\frac{\beta}{2}(\tilde{z}-\tilde{b})^{2}\left(\tilde{b}_{\tilde{x}}\tilde{u}_{\tilde{x}\tilde{x}}^{b}+(\tilde{b}_{\tilde{x}}\tilde{u}_{\tilde{x}}^{b}+
(\tilde{b}_{\tilde{x}}\tilde{u}^{b})_{\tilde{x}})_{\tilde{x}}\right)+
\frac{\beta}{3!}(\tilde{z}-\tilde{b})^{3}\tilde{u}_{\tilde{x}\tilde{x}\tilde{x}}^{b}+\mathcal{O}(\beta^{2})\ ,\label{eq:vvelocity}
\end{multline*}
then the incompressibility condition (\ref{eq:continuity1}) holds to $\mathcal{O}(\beta^{2})$.
Depth averaging (\ref{eq:hvelocity1}) gives
\begin{equation*}
 \tilde{u}^{b}=\bar{\tilde{u}}- \frac{\beta}{2}\tilde{h}\left(\tilde{b}_{\tilde{x}}\bar{\tilde{u}}_{\tilde{x}}+(\tilde{b}_{\tilde{x}}\bar{\tilde{u}})_{\tilde{x}}\right)+
 \frac{\beta}{6}\tilde{h}^{2}\bar{\tilde{u}}_{\tilde{x}\tilde{x}}+\mathcal{O}(\beta^{2},\alpha\beta^{2})\ .
\end{equation*}
Thus the horizontal velocity is
\begin{multline}  \label{eq:hvelocity2}
\tilde{u}(\tilde{x},\tilde{z},\tilde{t})=\bar{\tilde{u}}-\beta \left(\tilde{b}_{\tilde{x}}\bar{\tilde{u}}_{\tilde{x}}+(\tilde{b}_{\tilde{x}}\bar{\tilde{u}})_{\tilde{x}}\right)
\left(\frac{\tilde{h}}{2}-(\tilde{z}-\tilde{b})\right)\\
+\beta \left(\frac{\tilde{h}^{2}}{6}-\frac{1}{2}(\tilde{z}-\tilde{b})^{2}\right)\bar{\tilde{u}}_{\tilde{x}\tilde{x}}+\mathcal{O}(\beta^{2},\alpha\beta^{2})\ .
\end{multline}
Taking squares in equation ( \ref{eq:hvelocity2}) 
\begin{multline} \label{eq:hvelocitysqur}
\tilde{u}^{2}(\tilde{x},\tilde{z},\tilde{t})=\bar{\tilde{u}}^{2}-\beta\left(\tilde{b}_{\tilde{x}}\bar{\tilde{u}}_{\tilde{x}}\bar{\tilde{u}}+(\tilde{b}_{\tilde{x}}\bar{\tilde{u}})_{\tilde{x}}
\bar{\tilde{u}}\right)\left(\tilde{h}-2(\tilde{z}-\tilde{b})\right)+\\
\beta \left(\frac{\tilde{h}^{2}}{2}-(\tilde{z}-\tilde{b})^{2}\right)\bar{\tilde{u}} \bar{\tilde{u}}_{\tilde{x}\tilde{x}}+\mathcal{O}(\beta^{2},\alpha\beta^{2})\ .  
\end{multline}
Integrating equation (\ref{eq:hvelocitysqur}) from $\tilde{b}$ to $\alpha\tilde{\eta}$ and after some simplifications it follows that
\begin{equation}
 \int_{\tilde{b}}^{\alpha\tilde{\eta}}  \left(\tilde{u}^{2}-(\bar{\tilde{u}})^{2}\right)\,d\tilde{z}
= \mathcal{O}(\beta^{2},\alpha\beta^{2})\ ,
\end{equation}
and that
\begin{multline}\label{eq:gamma1}
\Gamma(\tilde{x},\tilde{z},\tilde{t})
 = (\tilde{z} - \tilde{b})\left[ 
      \alpha \bar{\tilde{u}}_{\tilde{x}}^{2}
    - \bar{\tilde{u}}_{\tilde{x}\tilde{t}}  
     - \alpha \bar{\tilde{u}} \bar{\tilde{u}}_{\tilde{x}\tilde{x}} \right]
+ \\
+ \tilde{b}_{\tilde{x}} (\bar{\tilde{u}}_{\tilde{t}} + \alpha \bar{\tilde{u}} \bar{\tilde{u}}_{\tilde{x}} )
+ \alpha \tilde{b}_{\tilde{x}\tilde{x}} \bar{\tilde{u}}^{2}
+\mathcal{O}(\beta,\alpha \beta)\ .
\end{multline}
Evaluating the integrals $\int_{\tilde{b}}^{\alpha \tilde{\eta}} \Gamma \, d \tilde{z}$
and $\int_{\tilde{b}}^{\alpha \tilde{\eta}} (\tilde{z} - \tilde{b}) \Gamma \, d \tilde{z}$
yields
\begin{equation}
\int_{\tilde{b}}^{\alpha \tilde{\eta}} \Gamma \, d \tilde{z}
= \frac{1}{2} \tilde{h} \tilde{\PP} + \tilde{h} \tilde{\QQ}\ ,
\end{equation}
and
\begin{equation}
\int_{\tilde{b}}^{\alpha \tilde{\eta}}  (\tilde{z} - \tilde{b}) \Gamma \, d \tilde{z}
= \frac{1}{3} \tilde{h}^2 \tilde{\PP} + \frac{1}{2}\tilde{h}^2 \tilde{\QQ}\ ,
\end{equation}
where
 \begin{equation}\label{eq:pp}
 \tilde{\PP} = \tilde{h} \left[ 
      \alpha \bar{\tilde{u}}_{\tilde{x}}^{2}
    - \bar{\tilde{u}}_{\tilde{x}\tilde{t}}  
     - \alpha \bar{\tilde{u}} \bar{\tilde{u}}_{\tilde{x}\tilde{x}} \right]\ ,
\end{equation}
and
\begin{equation}\label{eq:qq}
\tilde{\QQ} =  \tilde{b}_{\tilde{x}} \left( \bar{\tilde{u}}_{\tilde{t}} 
             + \alpha \bar{\tilde{u}} \bar{\tilde{u}}_{\tilde{x}} \right) + \tilde{b}_{\tilde{x}\tilde{x}}\bar{\tilde{u}}^2\ . 
\end{equation}
Finally we find the second equation of the system as
\begin{multline*}
\bar{\tilde{u}}_{\tilde{t}} + \alpha \bar{\tilde{u}}\bar{\tilde{u}}_{\tilde{x}} + \tilde{\eta}_{\tilde{x}}
+ \frac{\beta}{\tilde{h}}\frac{\partial}{\partial \tilde{x}}
    \left\{ \Big( \Sfrac{1}{3} \tilde{\PP}  + \Sfrac{1}{2} \tilde{\QQ} \Big) \tilde{h}^{2} \right\}
+ \beta \tilde{b}_{\tilde{x}} \Big( \Sfrac{1}{2} \tilde{\PP} +  \tilde{\QQ} \Big)
 = \mathcal{O}(\alpha\beta^{2})\ .
\end{multline*}

By setting the right-hand side equal to zero, and writing the variables in dimensional form the system reads 
\begin{subequations}\label{eq:DSerre}
\begin{align}
& \eta_{t}+[h\bar{u}]_{x}=0\ ,\label{eq:DSerre1} \\
& \bar{u}_{t} + \bar{u}\bar{u}_{x} + g\eta_{x}+
\frac{1}{h} \left[ h^{2} \big( \Sfrac{1}{3} \PP + \Sfrac{1}{2} \QQ \big) \right]_x
+ b_x \big( \Sfrac{1}{2} \PP +  \QQ \big) =0 \, ,
\label{eq:DSerre2}
\end{align}
\end{subequations}
where
$ \PP =  h \left[ \bar{u}_{x}^{2} - \bar{u}_{xt} - \bar{u} \bar{u}_{xx} \right]$ 
and
$ \QQ = b_x (\bar{u}_t + \bar{u}\bar{u}_x) + b_{xx} \bar{u}^2.$

In order to determine which terms 
should be kept for the velocity field at a certain order of approximation, 
the incompressibility condition (\ref{eq:continuity1}) can be used.
Then, the dimensional form of the water particle velocities 
at any location $(x,z)$ in the vertical plane become
\begin{subequations}\label{eq:velocts}
\begin{align}
 & u=\bar{u}+\left(\frac{h^{2}}{6}-\frac{z^{2}}{2}\right)\bar{u}_{xx}\ , \label{eq:velocts1} \\
 & v=-z\bar{u}_{x}\ . \label{eq:velocits2}
\end{align}
\end{subequations}
As it was mentioned before, system (\ref{eq:serre1a}) and (\ref{eq:serre1b}) 
reduces to the shallow water system when $\beta \rightarrow 0$ 
and to the classical Boussinesq system when $\beta \sim \alpha$.

An asymptotic expression for the pressure $\tilde{p}(\tilde{x},\tilde{z},\tilde{t})$ 
can be obtained by substituting formula (\ref{eq:gamma1}) into (\ref{eq:pressure2}). 
Such a formula was derived in \cite{PeliChoi} in the form
\begin{multline}\label{eq:usepress}
\tilde{p}(\tilde{x},\tilde{z},\tilde{t})
=\alpha \tilde{\eta}-\tilde{z} +
+ \frac{\alpha\beta}{2} \left[- \bar{\tilde{u}}_{\tilde{x}\tilde{t}}
- \alpha\bar{\tilde{u}}\bar{\tilde{u}}_{\tilde{x}\tilde{x}} 
+ \alpha \bar{\tilde{u}}_{\tilde{x}}^2\right] 
  \left( \tilde{h}^2 - ( \tilde{z} - \tilde{b} )^2 \right) 
\\
+ \alpha\beta \left( \alpha \tilde{b}_{\tilde{x}\tilde{x}} \bar{\tilde{u}}^2 
                      +  \alpha \tilde{b}_{\tilde{x}} \bar{\tilde{u}} \bar{\tilde{u}}_{\tilde{x}}
                      + \tilde{b}_{\tilde{x}} \bar{\tilde{u}}_{\tilde{t}} \right) (\alpha \tilde{\eta} - \tilde{z} )
+ \mathcal{O}(\alpha\beta^2)\ .
\end{multline} 

\section{Mechanical balance laws for the \acs{SGN} equations}\label{sec:laws}
In this section we derive the mechanical balance laws such as the mass, 
momentum and energy conservation for the \acs{SGN} equations extending the results related 
to some Boussinesq systems found in \cite{AlKa2012JNLS}. 
The balance laws consist of terms of the same asymptotic order as in the \acs{SGN} equations. 
We start with the conservation of mass.

\subsection{Mass balance}\label{ssec:massb}
We investigate the mass conservation properties of equations (\ref{eq:DSerre1}) and (\ref{eq:DSerre2}). 
Our starting point 
is the total mass of the fluid contained in a control volume of unit width, 
bounded by the lateral sides of the interval $[x_{1}, x_{2}]$, and by the free surface and the bottom. 
This mass is given by
\begin{equation}
 \mathcal{M}= \int_{x_{1}}^{x_{2}} \int_{b}^{\eta} \rho \, dz\, dx \, .
\end{equation}
According to the principle of mass conservation and the fact that there is no mass flux 
through the bottom or the free surface, 
mass conservation can be considered in terms of the flow variables as follows:
\begin{equation}
 \frac{d}{dt}\int_{x_{1}}^{x_{2}} \int_{b}^{\eta} \rho\, dz\, dx  
= \left[\int_{b}^{\eta} \rho u(x, z,t)\, dz \right]_{x_{2}}^{x_{1}}\, .
\end{equation}
In non-dimensional form this equation becomes
\begin{equation}
 \frac{d}{d\tilde{t}}\int_{\tilde{x}_{1}}^{\tilde{x}_{2}}\int_{\tilde{b}}^{\alpha\tilde{\eta}}\, d\tilde{z}\, d\tilde{x}=
 \alpha \left[\int_{\tilde{b}}^{\alpha\tilde{\eta}} \tilde{u}(\tilde{x},\tilde{z},\tilde{t})\, 
 d\tilde{z}\right]_{\tilde{x}_{2}}^{\tilde{x}_{1}}\ .
\end{equation}
Substituting the expression (\ref{eq:hvelocity2}) for $\tilde{u}$ and integrating with respect to $\tilde{z}$ yields
\begin{equation}\label{eq:MassB}
 \frac{d}{d\tilde{t}}\int_{\tilde{x}_{1}}^{\tilde{x}_{2}}\tilde{h}d\tilde{x}=\alpha\left[\bar{\tilde{u}}\tilde{h}%
 \right]_{\tilde{x}_{2}}^{\tilde{x}_{1}}+\mathcal{O}(\alpha\beta^2)\ , 
\end{equation}
where $\tilde{h}=\alpha\tilde{\eta}-\tilde{b}$ denotes the nondimensional total depth. Dividing (\ref{eq:MassB}) by $\tilde{x}_{2}-\tilde{x}_{1}$ and taking $\tilde{x}_2-\tilde{x}_1\rightarrow 0$ then 
the mass balance equation is written as
\begin{equation}\label{eq:nonDMassB}
\tilde{h}_{\tilde{t}} + (\alpha\bar{\tilde{u}}\tilde{h})_{\tilde{x}}=\mathcal{O}(\alpha\beta^2)\ . 
\end{equation}
Denoting the non-dimensional mass density by $\tilde{M}=\tilde{h}$ 
and the non-dimensional mass flux by $\tilde{q}_{M}=\alpha\bar{\tilde{u}}\tilde{h}$, then the mass balance is 
\begin{equation}
\frac{\partial \tilde{M}}{\partial{\tilde{t}}}
 + \frac{\partial{\tilde{q}_{M}}}{\partial{\tilde{x}}}= \mathcal{O}(\alpha\beta^2)\, .
\end{equation}
Using the scaling $M=\rho b_0\tilde{M}$ and $q_{M}=\rho b_0c_{0}\tilde{q}_{M}$
the dimensional forms of mass density and mass flux are $M=\rho h$ and $q_{M}=\rho\bar{u}h$
respectively. Then the dimensional form of the mass balance is obtained by discarding the right-hand side of the scaled mass balance equation and using the unscaled quantities:
\begin{equation}
\frac{\partial M}{\partial t} + \frac{\partial q_M }{\partial x} = 0\ .
\end{equation}
It is noted that the mass balance is satisfied exactly by the solutions of the \acs{SGN} system. 

The expressions for mass density and the mass flux do not depend on the shape of the
bottom topography, and in particular, they have the same form for both even and uneven beds.
The dimensional form of (\ref{eq:nonDMassB}) coincides with analogous formulas
of the shallow-water wave system and the classical Boussinesq system \cite{AlKa2012JNLS}. 
While this may be expected, it should be pointed out that in the case of other 
asymptotically equivalent systems, mass conservation may be satisfied only to the same 
order as the order of the equations, \cite{AlKa2012JNLS}.

\subsection{Momentum Balance}\label{sec:momentum}
The total horizontal momentum of a fluid of constant density $\rho$ contained in a control 
volume of the same type as in the previous section is 
\begin{equation}
 \mathcal{I}=\int_{x_{1}}^{x_{2}} \int_{b}^{\eta} \rho u \,dz~dx\, .
\end{equation}
Conservation of momentum implies that the rate of change of $\mathcal{I}$ 
is equal to the net influx of momentum through 
the boundaries plus the net force at the boundary of the control volume. 
Therefore, the conservation of momentum is written
\begin{equation*}
\frac{d}{dt}\int_{x_{1}}^{x_{2}} \int_{b}^{\eta} \rho u \,dz\, dx=  -\int_{x_{1}}^{x_{2}} pb_{x}~dx \\
+\left[\int_{b}^{\eta} \rho u^{2}(x, z)\,dz +
\int_{b}^{\eta}p dz\right]_{x_{2}}^{x_{1}} \ .
\end{equation*}
Non-dimensionalization of this expression leads to
\begin{equation*}
\alpha\frac{d}{d\tilde{t}}\int_{\tilde{x}_{1}}^{\tilde{x}_{2}}\int_{\tilde{b}}^{\alpha\tilde{\eta}}\tilde{u}\,d\tilde{z}\,d\tilde{x}= -
\int_{\tilde{x}_{1}}^{\tilde{x}_{2}} \tilde{P}_{b}\tilde{b}_{\tilde{x}}~d\tilde{x}\\
+\left[\alpha^{2}\int_{\tilde{b}}^{\alpha\tilde{\eta}}
\tilde{u}^{2}\, d\tilde{z}+\int_{\tilde{b}}^{\alpha\tilde{\eta}}\tilde{p}\,d\tilde{z}\right]_{\tilde{x}_{2}}^{\tilde{x}_{1}}\ , 
\end{equation*}
where $\tilde{P}_{b}$ denotes the pressure at the bottom $\tilde{P}_b=\tilde{h}
+ \frac{\alpha\beta}{2} [- \bar{\tilde{u}}_{\tilde{x}\tilde{t}}
- \alpha\bar{\tilde{u}}\bar{\tilde{u}}_{\tilde{x}\tilde{x}} 
+ \alpha \bar{\tilde{u}}_{\tilde{x}}^2 ] 
  \tilde{h}^2 + \alpha\beta ( \alpha \tilde{b}_{\tilde{x}\tilde{x}} \bar{\tilde{u}}^2 
                      +  \alpha \tilde{b}_{\tilde{x}} \bar{\tilde{u}} \bar{\tilde{u}}_{\tilde{x}}
                      + \tilde{b}_{\tilde{x}} \bar{\tilde{u}}_{\tilde{t}} ) \tilde{h}$.
Substituting the values of $\tilde{u}$ and $\tilde{p}$ 
from equations (\ref{eq:hvelocity2}) and (\ref{eq:pressure2}) 
and integrating with respect to $\tilde{z}$ yields
\begin{multline} \nonumber
\alpha\frac{d}{d\tilde{t}}\int_{\tilde{x}_{1}}^{\tilde{x}_{2}}\bar{\tilde{u}}\tilde{h}d\tilde{x} = - \int_{\tilde{x}_{1}}^{\tilde{x}_{2}} \tilde{P}_{b}\tilde{b}_{\tilde{x}}\, d\tilde{x}\\
+\left[\alpha^{2} \bar{\tilde{u}}^{2}\tilde{h} + \frac{\tilde{h}^{2}}{2}-
\frac{\alpha\beta}{3}\tilde{h}^{3}\Big(\bar{\tilde{u}}_{\tilde{x}\tilde{t}} +
\alpha\bar{\tilde{u}}\bar{\tilde{u}}_{\tilde{x}\tilde{x}}-\alpha(\bar{\tilde{u}}_{\tilde{x}})^{2}\Big)\right]_{\tilde{x}_{2}}^{\tilde{x}_{1}}+\\
\left[\frac{\alpha\beta}{2}\tilde{h}^{2}\Big(\alpha\tilde{b}_{\tilde{x}\tilde{x}}\bar{\tilde{u}}^{2}+ \tilde{b}_{\tilde{x}}( \alpha\bar{\tilde{u}}\bar{\tilde{u}}_{\tilde{x}}+
\bar{\tilde{u}}_{\tilde{t}}) \Big) \right]_{\tilde{x}_{2}}^{\tilde{x}_{1}} +
\mathcal{O}(\alpha\beta^{2})\ .
\end{multline}
Applying similar techniques used for the derivation of the mass balance equation we obtain the momentum balance equation in the form
\begin{multline}
\left(\alpha\bar{\tilde{u}}\tilde{h}\right)_{\tilde{t}} + \left(\alpha^{2}\bar{\tilde{u}}^{2}\tilde{h} +
\frac{\tilde{h}^{2}}{2} -
\frac{\alpha\beta}{3}\tilde{h}^{3}\Big(\bar{\tilde{u}}_{\tilde{x}\tilde{t}} + 
\alpha\bar{\tilde{u}}\bar{\tilde{u}}_{\tilde{x}\tilde{x}}-
\alpha(\bar{\tilde{u}}_{\tilde{x}})^{2}\Big)\right)_{\tilde{x}}+\\ 
\left(\frac{\alpha\beta}{2}\tilde{h}^{2}\Big(\alpha\tilde{b}_{\tilde{x}\tilde{x}}\bar{\tilde{u}}^{2}+
\tilde{b}_{\tilde{x}}(\alpha\bar{\tilde{u}}\bar{\tilde{u}}_{\tilde{x}}+
\bar{\tilde{u}}_{\tilde{t}})\Big)\right)_{\tilde{x}}= -\tilde{P}_b\tilde{b}_{\tilde{x}} + \mathcal{O}(\alpha\beta^{2})\ . \label{eq:MomB}
\end{multline}
If the non-dimensional momentum density is defined by
\begin{equation}
\tilde{I}=\alpha\bar{\tilde{u}}\tilde{h}\, 
\end{equation}
and the momentum flux plus pressure force is defined by
\begin{multline} \nonumber
\tilde{q}_{I}= \alpha^{2}\bar{\tilde{u}}^{2}\tilde{h} +
\frac{\tilde{h}^{2}}{2} -
\frac{\alpha\beta}{3}\tilde{h}^{3}(\bar{\tilde{u}}_{\tilde{x}\tilde{t}} + 
\alpha\bar{\tilde{u}}\bar{\tilde{u}}_{\tilde{x}\tilde{x}}-
\alpha(\bar{\tilde{u}}_{\tilde{x}})^{2})+\\ \frac{\alpha\beta}{2}\tilde{h}^{2}(\alpha\tilde{b}_{\tilde{x}\tilde{x}}\bar{\tilde{u}}^{2}+
\tilde{b}_{\tilde{x}}(\alpha\bar{\tilde{u}}\bar{\tilde{u}}_{\tilde{x}}+
\bar{\tilde{u}}_{\tilde{t}}))\,
\end{multline}
then the momentum balance equation can be written as
\begin{equation}
\frac{\partial{\tilde{I}}}{\partial{\tilde{t}}}+\frac{\partial{\tilde{q}_{I}}}{\partial{\tilde{x}}}
= -\tilde{P}_{b}\tilde{b}_{\tilde{x}}+ \mathcal{O}(\alpha\beta^{2})\ .
\end{equation}
Using the scaling $I=\rho c_{0}b_0\tilde{I}$ and $q_I=\rho c_{0}^{2}b_0\tilde{q_I}$, 
the dimensional forms of the momentum density and momentum flux per unit span are given by 
\begin{equation}
\label{I}
I=\rho\bar{u}h\ ,
\end{equation}
and
\begin{multline}
\label{qI}
 q_{I}=\rho \bar{u}^{2}h+\frac{\rho g}{2}h^{2}-\frac{\rho}{3}(\bar{u}_{xt} +
\bar{u}\bar{u}_{xx}-\bar{u}_{x}^{2})h^{3}+\\
+ \frac{\rho}{2} (b_{xx}\bar{u}^{2}+
b_{x}(\bar{u}\bar{u}_{x}+
\bar{u}_{t}))h^{2}\ ,
\end{multline}
respectively.
It turn out that the momentum conservation law is also satisfied exactly in the
context of the \acs{SGN} system. Indeed, if the momentum density 
is defined by (\ref{I}), the momentum flux plus pressure force is defined by (\ref{qI}),
and the pressure is defined by (\ref{eq:usepress}), then
solutions of the \acs{SGN} system also satisfy exactly the equation
\begin{equation}
\frac{\partial{I}}{\partial{t}}+\frac{\partial{q_{I}}}{\partial{x}}= - b_x p\ .
\end{equation}
Note that if the bottom $b=-b_0$ is horizontal, then the last equation 
is homogeneous and does not depend on the pressure $p$. 

Taking $\beta \rightarrow 0$ in the momentum balance equation (\ref{eq:MomB}), 
and using dimensional variables and horizontal bottom $b=-b_0$, 
the momentum density is unchanged, but the flux reduces to
\begin{equation}\label{eq:mom2}
q_{I}^{sw}=\rho \bar{u}^{2}h+\frac{\rho g}{2}h^{2}\ .
\end{equation}
Thus it is plain that both the momentum density $I$ and flux $q_{I}$ 
reduce correctly to the nonlinear shallow water approximation.
In the case $\beta \sim \alpha$ and a flat bottom, the
quantities for the momentum balance law are $I=\rho\bar{u}(b_{0}+\eta)$ and
$q_{I}=\rho b_{0} \bar{u}^{2}+\frac{\rho g}{2}h^{2}-\frac{\rho}{3}b_{0}^{3}\bar{u}_{xt}$,
which agree with the corresponding quantities of the classical Boussinesq system.


\subsection{Energy Balance}
The total mechanical energy inside a control volume can be written as the sum of the kinetic 
and potential energy as
\begin{equation}
\mathcal{E}= \int_{x_{1}}^{x_{2}} \int_{b}^{\eta} \left\{ \Sfrac{\rho}{2} (u^{2}+v^{2}) +
\rho g z \right\}\, dz\, dx\ .
\end{equation}
The conservation energy can be expressed as
\begin{multline}
\frac{d}{dt} \int_{x_{1}}^{x_{2}} \int_{b}^{\eta} \left\{ \Sfrac{\rho}{2} (u^{2}+v^{2}) + 
 \rho g z  \right\}\,dz\, dx =\\
 = \left[\int_{b}^{\eta} \left\{ \left(\Sfrac{\rho}{2}  (u^{2}+v^{2}) + \rho g z \right)u\, + 
 u P \right\} \,dz\right]_{x_{2}}^{x_{1}}\ , \label{eq:EnergyB}
\end{multline}
and in non-dimensional variables as
\begin{multline}
 \frac{d}{d\tilde{t}}\int_{\tilde{x}_{1}}^{\tilde{x}_{2}}\int_{\tilde{b}}^{\alpha\tilde{\eta}} \left\{ \Sfrac{\alpha^{2}}{2}(\tilde{u}^{2}+
 \beta\tilde{v}^{2}) \,
 +  \tilde{z} \right\} \,d\tilde{z}\, d\tilde{x} = \\
 \alpha \left[ \int_{\tilde{b}}^{\alpha\tilde{\eta}}\left\{ 
 \Sfrac{\alpha^{2}}{2}(\tilde{u}^{3}+\beta\tilde{v}^{2}\tilde{u})+
 \tilde{z}\tilde{u} \,  + 
 \tilde{p}\tilde{u} \right\} \,d\tilde{z}\right]_{\tilde{x}_{2}}^{\tilde{x}_{1}}. \label{eq:nonEB1}
\end{multline}
By substituting the expressions (\ref{eq:lowhvel1}), (\ref{eq:lowvvel1}) and (\ref{eq:usepress}) 
for $\tilde{u}$, $\tilde{v}$ and $\tilde{p}$ respectively, 
the energy balance equation takes the form
\begin{multline}
 \frac{d}{d\tilde{t}}\int_{\tilde{x}_{1}}^{\tilde{x}_{2}}\left(\frac{\alpha^{2}}{2}\Big(\bar{\tilde{u}}^{2}+\beta {\tilde{b}_{\tilde{x}}}^{2}\bar{\tilde{u}}^{2}\Big)\tilde{h}
 -\frac{\alpha^{2}\beta}{2}\tilde{b}_{\tilde{x}}\tilde{h}^{2}\bar{\tilde{u}}\bar{\tilde{u}}_{\tilde{x}}+ \right. \\  \left.\frac{\alpha^{2}\beta}{6}\tilde{h}^{3}
 \bar{\tilde{u}}_{\tilde{x}}^{2}+\frac{\tilde{h}^{2}}{2}+\tilde{b}\tilde{h}\right)d\tilde{x}  =\left[\frac{\alpha^{3}}{2}\bar{\tilde{u}}^{3}\Big(1+\beta{\tilde{b}_{\tilde{x}}}^{2}\Big)
 \tilde{h}+ \frac{\alpha}{2}
 \tilde{h}^{2}\bar{\tilde{u}}+\right. \\  \left. \alpha\tilde{b}\bar{\tilde{u}}\tilde{h}-\frac{\alpha^{3}\beta}{2}\tilde{b}_{\tilde{x}}\bar{\tilde{u}}_{\tilde{x}}\bar{\tilde{u}}^{2}
 \tilde{h}^{2}+\frac{\alpha^{3}\beta}{6}\bar{\tilde{u}}\bar{\tilde{u}}_{\tilde{x}}^{2}\tilde{h}^{3}\right]_{\tilde{x}_{2}}^{\tilde{x}_{1}} +
 \\ \left[\frac{\alpha}{2}\bar{\tilde{u}}\tilde{h}^{2}-\frac{\alpha^{2}\beta}{3}
 \tilde{h}^{3}\bar{\tilde{u}}\Big(\bar{\tilde{u}}_{\tilde{x}\tilde{t}}+\alpha\bar{\tilde{u}}\bar{\tilde{u}}_{\tilde{x}\tilde{x}} - \alpha\bar{\tilde{u}}_{\tilde{x}}^{2}\Big)\right. \\ 
  \left. -\frac{\alpha^{2}\beta}{2}\Big(\alpha\tilde{b}_{\tilde{x}\tilde{x}}\bar{\tilde{u}}^{2}+
\alpha\tilde{b}_{\tilde{x}}(\bar{\tilde{u}}\bar{\tilde{u}}_{\tilde{x}}+
\bar{\tilde{u}}_{\tilde{t}}) \Big)\tilde{h}^{2}\right]_{\tilde{x}_{2}}^{\tilde{x}_{1}}
+ \mathcal{O}(\alpha\beta^{2}). \label{eq:nonEB2}
\end{multline}
The differential form of the energy balance equation is given by
\begin{multline}
\left(\frac{\alpha^{2}}{2}\Big(\bar{\tilde{u}}^{2}+\beta {\tilde{b}_{\tilde{x}}}^{2}\bar{\tilde{u}}^{2}\Big)\tilde{h}
 -\frac{\alpha^{2}\beta}{2}\tilde{b}_{\tilde{x}}\tilde{h}^{2}\bar{\tilde{u}}\bar{\tilde{u}}_{\tilde{x}}+\frac{\alpha^{2}\beta}{6}\tilde{h}^{3}\bar{\tilde{u}}_{\tilde{x}}^{2}+
 \frac{\tilde{h}^{2}}{2}+\tilde{b}\tilde{h}\right)_{\tilde{t}}\\ 
 +\left(\frac{\alpha^{3}}{2}\bar{\tilde{u}}^{3}\tilde{h}+\frac{\alpha^{3}\beta}{3}\tilde{b}_{\tilde{x}}^{2}\bar{\tilde{u}}^{3}+\alpha\bar{\tilde{u}}\tilde{h}^{2}+\alpha\tilde{b}\bar{\tilde{u}}
 \tilde{h}- \frac{\alpha^{2}\beta}{3}
 \tilde{h}^{3}\bar{\tilde{u}}\Big(\bar{\tilde{u}}_{\tilde{x}\tilde{t}}+  \alpha\bar{\tilde{u}}
 \bar{\tilde{u}}_{\tilde{x}\tilde{x}}\right. \\ \left.
 -\frac{3}{2}\alpha\bar{\tilde{u}}_{\tilde{x}}^{2}\Big) - \frac{\alpha^{3}\beta}{2}\tilde{b}_{\tilde{x}}\bar{\tilde{u}}_{\tilde{x}}\bar{\tilde{u}}^{2}
 \tilde{h}^{2}\right)_{\tilde{x}}\\-\left(\frac{\alpha^{2}\beta}{2}\tilde{h}^{2}\Big(\alpha\tilde{b}_{\tilde{x}\tilde{x}}\bar{\tilde{u}}^{2}+
\tilde{b}_{\tilde{x}}(\alpha\bar{\tilde{u}}\bar{\tilde{u}}_{\tilde{x}}+\bar{\tilde{u}}_{\tilde{t}})\Big)\right)_{\tilde{x}}
=\mathcal{O}(\alpha\beta^{2})\ .
\end{multline}
Considering the appropriate terms in the energy density and flux in (\ref{eq:nonEB1}) which are of order zero or one in the differential
energy balance (\ref{eq:nonEB2}), we find that the non-dimensional energy density is 
\begin{multline}
 \tilde{E} = \frac{\alpha^{2}}{2}(\bar{\tilde{u}}^{2}
 + \beta {\tilde{b}_{\tilde{x}}}^{2}\bar{\tilde{u}}^{2})\tilde{h}
 - \frac{\alpha^{2}\beta}{2}\tilde{b}_{\tilde{x}}
\tilde{h}^{2}\bar{\tilde{u}}\bar{\tilde{u}}_{\tilde{x}} 
  + \frac{\alpha^{2}\beta}{6}\tilde{h}^{3}\bar{\tilde{u}}_{\tilde{x}}^{2}
  + \frac{\tilde{h}^{2}}{2}+\tilde{b}\tilde{h}\, ,
\end{multline}
while the non-dimensional energy flux plus the work rate due to pressure forces is written as
\begin{multline}
 \tilde{q}_{E}
= \frac{\alpha^{3}}{2}\bar{\tilde{u}}^{3}\tilde{h}+\frac{\alpha^{3}\beta}{2}\tilde{b}_{\tilde{x}}^{2}\bar{\tilde{u}}^{3}+\alpha\tilde{b} \bar{\tilde{u}}\tilde{h}+\alpha\bar{\tilde{u}}\tilde{h}^{2} \\
- \frac{\alpha^{2}\beta}{3} \tilde{h}^{3}\bar{\tilde{u}}\Big(\bar{\tilde{u}}_{\tilde{x}\tilde{t}}
+ \alpha\bar{\tilde{u}} \bar{\tilde{u}}_{\tilde{x}\tilde{x}}
- \frac{3}{2}\alpha\bar{\tilde{u}}_{\tilde{x}}^{2}\Big)  
- \frac{\alpha^{3}\beta}{2}\tilde{b}_{\tilde{x}}\bar{\tilde{u}}_{\tilde{x}}\bar{\tilde{u}}^{2}
 \tilde{h}^{2} \\ 
-\frac{\alpha^{2}\beta}{2}\tilde{h}^{2}\bar{\tilde{u}}\Big(\alpha\tilde{b}_{\tilde{x}\tilde{x}}\bar{\tilde{u}}^{2} 
+ \tilde{b}_{\tilde{x}}(\alpha \bar{\tilde{u}}\bar{\tilde{u}}_{\tilde{x}}
+ \bar{\tilde{u}}_{\tilde{t}})\Big) .
\end{multline}
With these definitions, the energy balance is
\begin{equation}\label{eq:Ebal}
 \frac{\partial{\tilde{E}}}{\partial{\tilde{t}}}+\frac{\partial{\tilde{q}_{E}}}{\partial{\tilde{x}}}=
 \mathcal{O}(\alpha\beta^{2})\ .
\end{equation}
Using the scaling $E=\rho c_{0}^{2}b_0\tilde{E}$ and $q_{E}= \rho c_{0}^{3}b_0\tilde{q_{E}}$, 
the dimensional form of energy density per unit span in the transverse direction
is given as the sum of the kinetic and the potential energy by 
\begin{equation}\label{eq:endens}
E= \underbrace{\frac{\rho }{2}\bar{u}^{2}(1+b_{x}^{2})h
- \frac{\rho }{2}\bar{u}\bar{u}_{x}b_{x}h^{2}
+ \frac{\rho}{6}\bar{u}_{x}^{2}h^{3}}_{E_k}+\underbrace{\frac{\rho g}{2}h^{2}+\rho g b h}_{E_p}\, ,
\end{equation}
and the dimensional form of energy flux plus work rate due to the pressure force is given by
\begin{multline}
q_{E}=\rho g\bar{u}(h^{2}+bh)+\frac{\rho }{2} \bar{u}^{3}h(1+b_{x}^{2})-\frac{\rho}{3}h^{3}\bar{u}\Big(\bar{u}_{xt}+
 \bar{u}\bar{u}_{xx}-\frac{3}{2}\bar{u}^{2}_{x}\Big)\\-
 \frac{\rho }{2} \bar{u}^{2}\bar{u}_{x}b_{x}h^{2}+\frac{\rho }{2}\bar{u}h^{2}\Big(b_{xx}\bar{u}^{2}+b_{x}(\bar{u}\bar{u}_{x}+
 \bar{u}_{t})\Big) \, .
\end{multline}
For a horizontal bed, it is more convenient to
normalize the potential energy of a fluid particle to be zero at the bottom.
If this is done, then the dimensional 
forms of energy density and energy flux plus work rate due to pressure forces are given by
\begin{equation}
E = \frac{\rho g}{2}h^{2}
  + \frac{\rho }{2} h\bar{u}^{2}+\frac{\rho}{6}h^{3}\bar{u}_{x}^{2}\ ,
\label{EGN}
\end{equation}
and
\begin{equation}
 q_{E} = \rho g\bar{u}h^{2}+\frac{\rho }{2} \bar{u}^{3}h
      - \frac{\rho}{3}h^{3}\bar{u} \Big( \bar{u}_{xt} + \bar{u}\bar{u}_{xx}
      - \frac{3}{2}\bar{u}^{2}_{x} \Big) \ ,
\label{qEGN}
\end{equation}
respectively. Note that as $\beta\rightarrow 0$ in the equation (\ref{eq:nonEB2}), 
the energy balance reduces to the shallow-water energy 
conservation with 
\begin{equation}
E^{sw}=\frac{\rho g}{2}\Big(b_0^2 + 2b_0\eta+\eta^{2}\Big) + \frac{\rho}{2} h\bar{u}^{2}
\label{Esw}
\end{equation}
and 
\begin{equation}
q_{E}^{sw}=\rho g h^{2}\bar{u} + \frac{\rho}{2}  h\bar{u}^{3}.
\label{qEsw}
\end{equation}
In addition, in the case $\alpha \sim \beta$, 
the energy balance reduces correctly to the case of the classical Boussinesq system,
with $E = \frac{\rho g}{2}(b_0^2 + 2b_0\eta+\eta^{2}) + \frac{\rho}{2} b_0\bar{u}^{2}$ and
$q_E = \rho g (b_0^{2}+2\eta b_0)\bar{u}$.

It is worth noting that the conservation of the asymptotic approximation 
to the total energy with nontrivial bathymetry in the fully nonlinear regime 
is satisfied by the solutions of the \acs{SGN} equations exactly.
This can be seen by performing lengthy computations using formal integrations
by parts, or by recognizing that the total energy $\mathcal{E} = \int E \, dx$
differs from the Hamiltonian (\ref{eq:ham}) by a term $\frac{g}{2}\int b^2 \, dx$ which
is constant.


\section{Applications}\label{sec:applications}
\subsection{Evolution of undular bores}
In free surface flow, the transition between two states of different flow depth
is called a hydraulic jump if the transition region is stationary, and a bore if
it is moving. 
Bores are routinely generated by tidal forces in several rivers around the world,
and may also be generated in wavetank experiments \cite{Favre1935,chanson2009}.

The experimental studies of \cite{Favre1935} show that when the ratio between 
the difference in flow depths to the undisturbed water depth is smaller than $0.28$, 
then the bore will feature oscillations in the downstream part. 
If this ratio is greater than approximately $0.75$, 
then a so-called turbulent bore ensues.
If the ratio is between $0.28$ and $0.75$, the bore will be partially turbulent, 
but will also feature some oscillations. 
The bore strength can also be expressed in terms of the
Froude number $Fr=\sqrt{[(2h_1/h_0+1)^2-1]/8}$,
and more recent studies, such as \cite{chanson2009} 
have found that
when $Fr\geq 1.4$ approximately, the bore consists of a steep front, 
while undulations are growing at the bore front only in the near-critical state $Fr\approx 1$. 
The different shapes and a transition
from the subcritical to the supercritical regime is described in \cite{TBMCL2011},
and in \cite{TBMCL2012} an empirical critical value $Fr_{crit}=1.3$ 
is suggested in order to determine the breaking of an undular bore. 
However the exact characterization of the transition between these states still remains unclear. 

Some of the divergence in the results on the critical bore strength might be explained
by the observation that one single nondimensional number may not be sufficient to
classify all bores. For example, in \cite{RG2013}, a hyperbolic shear-flow
model is suggested which allows the classification of bores with an additional
parameter depending on the strength of the developing shear flow near the bore front.

The connection between the initial bore strength and the ensuing highest undulation is fairly well understood.
Using Whitham modulation theory \cite{Whitham1965}, it can be shown that if viscosity is neglected,
the amplitude of the leading wave behind
the bore front is exactly twice the initial ratio of flow depths \cite{Whitham1999, Grimshaw2007}.
This result agrees well with experimental findings. For example, the amplitude of the leading wave found 
experimentally in \cite{Favre1935} was 2.06 times the initial amplitude ratio.

In this section we present a numerical study 
of the energy balance of undular bores for the \acs{SGN} equations. 
It is well known that a sharp transition in both flow depth and flow velocity
which conserves both mass and momentum necessitates a loss of energy across
the front. The standard argument essentially relies on an inviscid shallow-water
theory and the examination of exact weak solutions of the shallow-water equations
\cite{Rayleigh1914}.

Given these assumptions, it is natural to explain the energy loss across the bore front
by pointing to the physical effects neglected in the shallow-water theory, such as
viscosity, frequency dispersion, and turbulent flow.
Indeed, in strong bores, turbulent dissipation accounts for the lion's share of energy
dissipation, and a long-wave model can only give a first approximation of the dynamics.
Most of the work investigating the energy loss has focused on weak undular bores,
where long-wave models can be expected to yield an accurate description of the flow.
The loss of energy in weak bores has been explained by the creation of oscillations
in the free surface behind the front, but it was noted in \cite{BenjaminLighthill1954}
that an additional dissipation mechanism is needed. In \cite{Sturtevant1965}, the bottom
boundary layer was invoked to explain this required additional energy loss, but
it was noted in \cite{AK2010, AK2012} that invoking frictional effects to explain
the energy loss experienced by a conservative system was not consistent.

However, as already mentioned, there was a slight technical problem  
in the analysis of \cite{AK2012}, since the energy functional
\begin{equation}\label{eq:enbous}
\mathcal{E}_{Bous}
= \frac{1}{2} \int_{\tilde{x}_{1}}^{\tilde{x}_{2}} \left[\alpha^2\tilde{w}^2 \tilde{h} 
                                     + \Sfrac{\alpha^2\beta}{3} \left(\tilde{w}\tilde{w}_{\tilde{x}\tilde{x}}
                                     + \tilde{w}_{\tilde{x}}^2 \right) + \tilde{h}^2\right]~ d\tilde{x}\ ,
\end{equation}
used in that work could not be obtained in the framework of the
asymptotically correct mechanical balance laws derived in \cite{AlKa2012JNLS}. 
Indeed, the expressions for the energy and energy flux associated to the
Boussinesq system which were derived in \cite{AlKa2012JNLS} are
\begin{equation*}
\tilde{E} =\frac{1}{2} + \alpha \tilde{\eta} + \frac{\alpha^2}{2}\tilde{\eta}^2  
                                 + \frac{\alpha^2}{2}\tilde{w}^2 
          = \frac{\alpha}{2} \tilde{h}^2 +  \frac{\alpha^2}{2}\tilde{w}^2\ ,
\end{equation*}
and the non-dimensional energy flux (corrected for the work rate due to pressure forces) as
\begin{multline*}
\tilde{q}_E 
= \alpha\tilde{w} + 2 \alpha^2 \tilde{w} \tilde{\eta} 
            +  \frac{\alpha\beta}{2} \left( \theta^2 - \frac{1}{3} \right) \tilde{w}_{\tilde{x}\tilde{x}}
 =\alpha\tilde{w} + 2 \alpha^2 \tilde{w} \tilde{\eta} 
            +  \frac{\alpha\beta}{6} \tilde{w}_{\tilde{x}\tilde{x}}\ ,
\end{multline*}
where $\tilde{w}$ is the nondimensional horizontal velocity component at height $\theta = b_0\sqrt{2/3}$
in the water column.
It is apparent that as $\beta \to 0$, these expressions do not reduce correctly 
to the corresponding expressions of the shallow-water theory. 
However, since the expressions (\ref{EGN}) and (\ref{qEGN}) 
do reduce to the correct shallow-water equivalents, 
the analysis of the energy loss in the undular bore can be made 
precise in the context of the \acs{SGN} system.

The numerical method that was used to perform the numerical simulations in this paper
is detailed in \ref{app:A}. 
It is also noted that for simplicity's sake we consider the water density $\rho=1~kg/m^3$.
The numerical experiments require initial data.
An initial surface condition that triggers the generation of undular bores is 
$$h(x,0)=h_0+\frac{1}{2}(h_1-h_0)\tanh(\kappa x)\ ,$$
where $\kappa$ is the parameter that determines the steepness of the undular bore. 
Here we take $\kappa=1/2$. In order to generate a simple undular bore,
\ie a wave that propagates mainly in one direction, 
we consider an initial flow given by the following velocity profile:
\begin{multline*}
u(x,0)
=\frac{\delta h}{h_1}\left(\frac{g}{2h_0}\left(2h_0^2+3(\delta h)h_0
  +(\delta h)^2 \right)\right)^{1/2}
  \times \left(1-\tanh(\kappa x) \right)\ ,
\end{multline*}
where $\delta h=h_1-h_0$.
One may envision other numerical methods to create an undular bore, such as the addition
of a line source in the upstream part, such as used in \cite{Tyvand2009}. Nevertheless,
the initial conditions described above were sufficient for our purposes.

\begin{table}[ht]
\caption{Energy conservation}
\begin{center}
\begin{tabular}{cccc}
$h_1/h_0$ & $Fr$ &  $q_E(x_1)-q_E(x_2) $
&  $d\mathcal{E}/dt$  \\
 \hline
 $1.1$ & $1.07$ & $3.6481059$ & $3.6481059$  \\
 $1.2$ & $1.15$ & $8.6017456$ & $8.6017456$  \\
 $1.3$ & $1.22$ & $15.100378$ & $15.100378$  \\
 $1.4$ & $1.30$ & $23.394470$ & $23.394470$  \\
 $1.5$ & $1.37$ & $33.746103$ & $33.746103$  \\
 $1.6$ & $1.44$ & $46.429376$ & $46.429376$  \\ 
 $1.7$ & $1.51$ & $61.730669$ & $61.730669$   \\
 \hline
\end{tabular}
\end{center}
\label{tab:tav1}
\end{table}
\begin{table}[ht]
\caption{Momentum conservation}
\begin{center}
\begin{tabular}{cccc}
$h_1/h_0$ & $Fr$ & $q_I(x_1)-q_I(x_2)$  &  $d\mathcal{I}/dt$  \\
 \hline
 $1.1$ & $1.07$ & $1.1330550$ & $1.1330549$   \\
 $1.2$ & $1.15$ & $2.5898340$ & $2.5898399$   \\
 $1.3$ & $1.22$ & $4.3997850$ & $4.3997849$   \\
 $1.4$ & $1.30$ & $6.5923199$ & $6.5923198$   \\
 $1.5$ & $1.37$ & $9.1968749$ & $9.1968748$   \\
 $1.6$ & $1.44$ & $12.242880$ & $12.242879$   \\ 
 $1.7$ & $1.51$ & $15.759764$ & $15.759764$    \\
 \hline
\end{tabular}
\end{center}
\label{tab:tav2}
\end{table}
First we present the computation of the energy budget in an undular bore
for various bore strengths.
We consider the control volume $[x_1,x_2]$, where $x_1$ is far to the left of the bore
front, and $x_2$ is far to to the right. In Table \ref{tab:tav1} 
the bore strength is shown in the first column, and the corresponding Froude number
is shown in the second column.
Taking $h_0=1$ and $h_1$ between $1.1$ and $1.7$ we monitor
the energy flux and work rate due to the pressure force, 
given by $q_E(x_1)-q_E(x_2)$ as defined in (\ref{qEGN}),
and these values are shown in the third column of the table.
We also monitor the gain in energy in the control interval as given by
$\mathcal{E}(t)=\int_{x_1}^{x_2} E~dx$.
These values are shown in the fourth column.
The particular figures shown in the table are for $T=30$, but the values are
nearly constant over time. 
It is apparent from the table that energy conservation holds to at least eight digits,
even for large bore strengths. These numbers confirm our previous finding that the energy
is exactly conserved in the \acs{SGN} model, and also validates the implementation of
the numerical method. In addition, these results confirm our claim that no dissipation
mechanism is necessary to explain the energy loss in an undular bore.

As noted in the previous section, the expression (\ref{qEGN}) for the energy flux 
and work rate due to pressure forces reduces to the corresponding formula for the 
shallow-water theory in the case of very long waves. Since $x_1$ and $x_2$ are relatively
far from the bore front, shallow-water theory should be valid at these points. 
Therefore, the usual formula for the energy loss in an undular bore in the shallow-water
theory is valid:
\begin{multline}
\label{EnergyEstimate}
\frac{d E^{sw}}{dt} + q_E^{sw}(x_2,t)-q_E^{sw}(x_1,t)
=-\frac{\rho}{4} (h_1-h_0)^3 \sqrt{ \half g^3 \left( \Sfrac{1}{h_0}+ \Sfrac{1}{h_1} \right) }\ .
\end{multline}
Since there is no energy loss in a dispersive system, one may conclude
that the excess energy is fed into oscillations of the free surface,
and the formula (\ref{EnergyEstimate}) furnishes an estimate of the
amount of energy which is residing in the oscillatory motion.

A similar study can be performed on the momentum balance.
Momentum gain in the control interval is given by the momentum flux 
through the lateral boundaries and the pressure force as
$q_I(x_1)-q_I(x_2)$, with $q_I$ given in (\ref{qI}) up to $T=30$. 
Table \ref{tab:tav2} presents the momentum rates. 
As in the case of the energy, the corresponding values agree to about eight digits.
\begin{figure}[ht]
  \centering
  \includegraphics[width=\columnwidth]{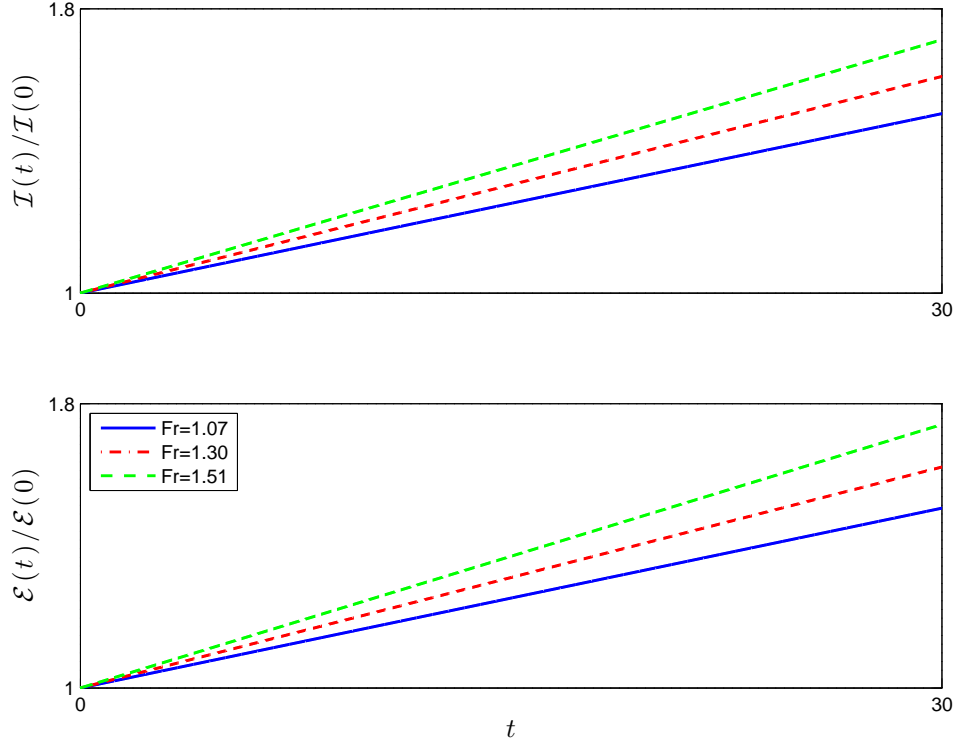}
  \caption{The momentum and the energy of the undular bore for $Fr=1.07$, $1.30$ and $1.51$.}
  \label{fig:figure3}
\end{figure}
In Figure \ref{fig:figure3}, we present the normalized values 
$\mathcal{I}(t)/\mathcal{I}(0)$ of the momentum and $\mathcal{E}(t)/\mathcal{E}(0)$ 
of the total energy for the values of the Froude number $Fr=1.07$, $1.30$ and $1.51$. 
The slopes of the lines can be found in Tables \ref{tab:tav1} and \ref{tab:tav2}.

\begin{figure}
  \centering
  \includegraphics[width=\columnwidth]{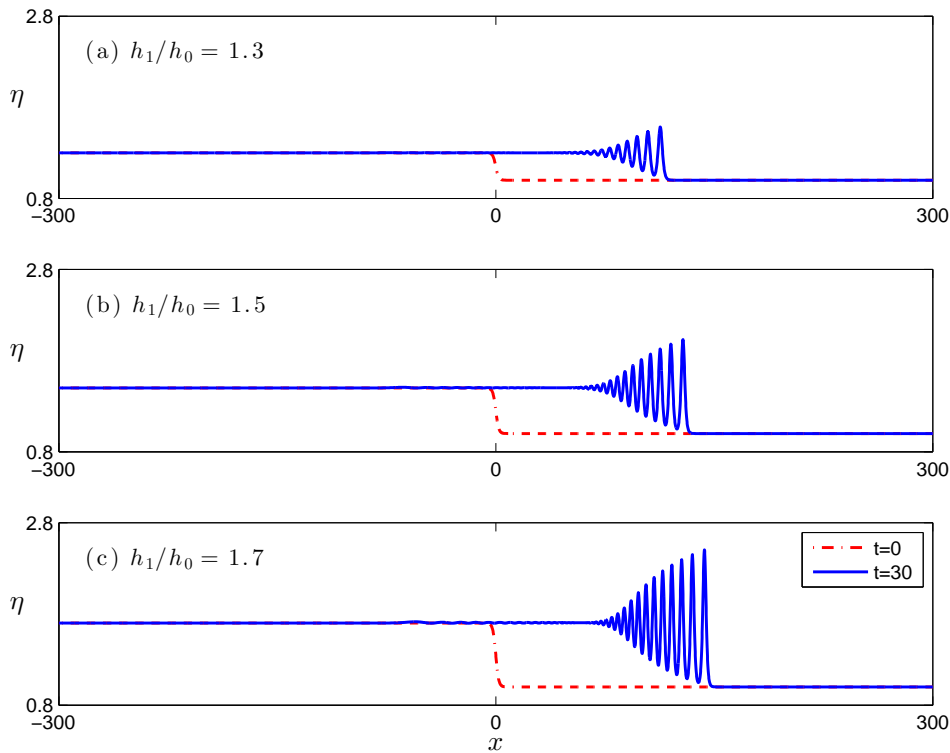}
  \caption{Undular bores profiles for various $Fr$ values.}
  \label{fig:figure1}
\end{figure}

Figure \ref{fig:figure1} shows the profiles of the undular bores 
generated when $h_1/h_0=1.3$, $1.5$ and $1.7$. 
From these figures we observe that as the Froude number $Fr$ increases, 
the peak amplitude of the leading wave becomes larger, 
and the shape of the wave envelope is changing. 
For example the shape of the wave envelope of Figure \ref{fig:figure1} (a) 
can be described by a quadratic function (wineglass shape) 
while the shape of the wave envelope of Figure \ref{fig:figure1} (c) 
can be described by a square-root function (martini glass shape). 
For the various shapes of the undular bores we refer to \cite {Hoefer2014}.   

\subsection{Shoaling of solitary waves}
In this section we study the conservation of energy in the case of a nonuniform bathymetry. 
Specifically, we consider the experiments proposed in \cite{GSSV1994,GSS1997} 
related to the shoaling of solitary waves on a beach of slope $1:35$. 
The shoaling of solitary waves has been studied theoretically and experimentally in many works,
such as in \cite{GSSV1994,GSS1997,Syn1991,SynSkj1993}. 
Next, we study the shoaling of solitary waves with normalized  
amplitude $A=0.1$, $0.15$, $0.2$ and $0.25$ in the domain $[-100,34]$. 
In the numerical experiments we take $\Delta x=0.05$ while we translate the solitary waves such 
that the peak amplitude is achieved at $x=-20.1171$ while the bottom is described by the function 
$$b(x)=\left\{\begin{array}{ll} -1, & x\leq 0\\ -1+x/35, & x> 0 \end{array}\right.\ ,$$ 
but modified appropriately around $x=0$ so as to be smooth enough and to satisfy 
the regularity requirements of the model. 

\begin{figure}
  \centering
  \includegraphics[width=\columnwidth]{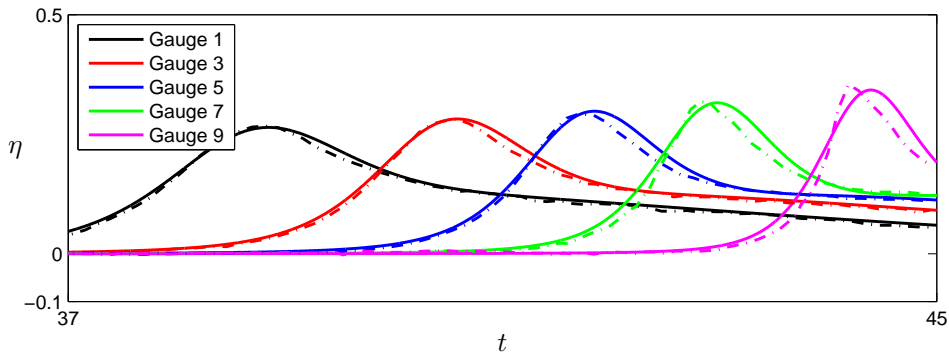}
  \caption{Comparison of the numerical solution and the experimental data on wave gauges of \cite{GSS1997}. ----: Numerical solution; $-\cdot-$: Experimental data.}
  \label{fig:figure5}
\end{figure}

A comparison between the experimental results on shoaling waves from \cite{GSSV1994},
and the shoaling solitary waves computed with numerically approximation of
(\ref{eq:serre1}) and (\ref{eq:serre2}) is presented in Figure \ref{fig:figure5}.
Overall, we observe a very good agreement between the numerical results and the experimental data.

Table \ref{tab:table4} presents the conserved values of the 
total energy $\mathcal{E}$ and of the Hamiltonian $\mathcal{H}$ for $t\in [0,45]$
for the computations shown in Figure \ref{fig:figure5}.
We observe that the energy is conserved with  more than ten decimal digits. 
Due to the small values of $\Delta x$ and $\Delta t$ no energy dissipation 
can be observed verifying the efficacy of the numerical method. 
\begin{table}[ht]
\caption{Conserved values of energy (in Joules) and Hamiltonian for shoaling of solitary waves on a plane beach of slope $1:35$.}
\begin{center}
\begin{tabular}{ccc}
\hline
$A$ & $\mathcal{E}$ & $\mathcal{H}$\\
\hline
 $0.10$ &  $62.4704607870$  & $0.05202930490$ \\
 $0.15$ &  $62.7102258381$  & $0.09856973753$ \\
 $0.20$ &  $62.9401348199$  & $0.15627417412$ \\
 $0.25$ &  $63.1680884219$  & $0.22460417742$ \\
 \hline
\end{tabular}
\end{center}
\label{tab:table4}
\end{table}

\begin{figure}
  \centering
  \includegraphics[width=\columnwidth]{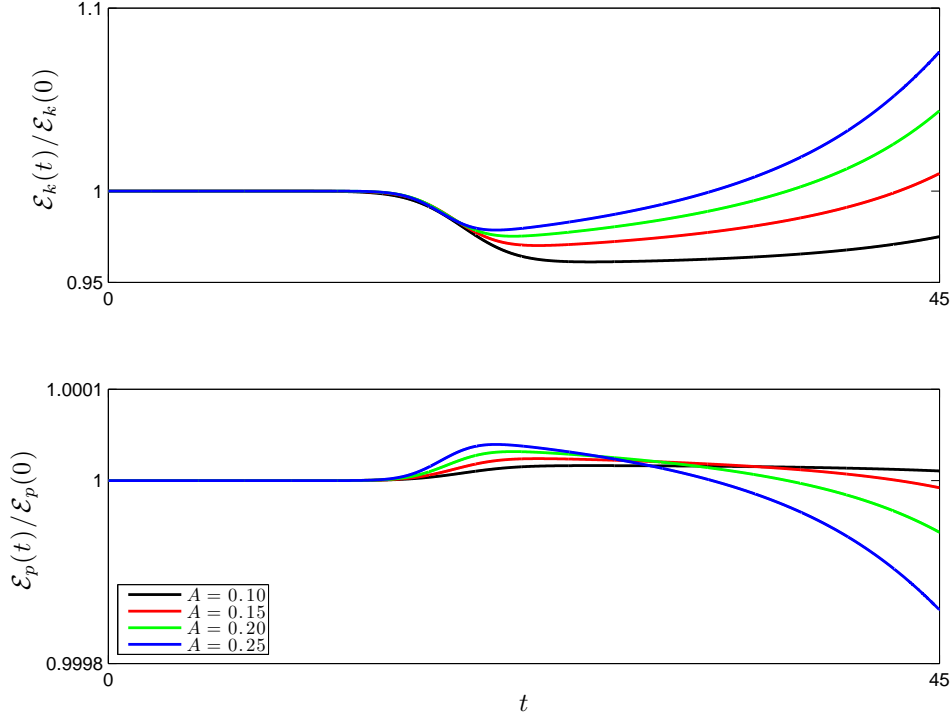}
  \caption{Normalized Kinetic and Potential energy for shoaling of solitary waves on a plane beach of slope $1:35$.}
  \label{fig:figure4}
\end{figure}

Although the total energy is conserved the kinetic and the potential energy are not constant with time. 
Figure \ref{fig:figure4} presents the normalized kinetic energy $\mathcal{E}_k(t)/\mathcal{E}_k(0)$ 
and normalized potential energy $\mathcal{E}_p(t)/\mathcal{E}_p(0)$ evaluated in the spatial interval $[-100,34]$. 
As can be seen in Figure \ref{fig:figure4} the kinetic energy is decreasing 
at the early stages of shoaling due to the slight decrease in the wave speed 
while the potential energy is initially increasing due to the increase of the wave height.
At later stages of the shoaling, the kinetic energy increases again, 
due to the increase in particle velocities,
and the potential energy decreases again, due to the rising bottom, and narrowing wave peak.
Nevertheless, the total energy is constant over time. 

\section{Summary and Conclusions}
We have detailed the derivation of mechanical balance laws for the 
\acs{SGN} equations in the case of a horizontal bed and also in the case of varying bathymetry. 
The mechanical balance laws derived here, including the mass,
momentum and energy balance laws, are valid to the same asymptotic order as the \acs{SGN} system, 
providing a firm link between conservation laws associated to the governing \acs{SGN} equations,
and the above mechanical quantities.
Finally, applications to the energy budget of undular bores and the development of potential and kinetic
energy in shoaling solitary waves have been presented. In particular, it has been shown that
the energy loss in undular bores is fully compensated for by the development of surface oscillations,
since the energy in the \acs{SGN} with a flat bottom is exactly conserved. Indeed, exact conservation of
energy to near machine precision was observed in our numerical computations, and this gave an additional
check on the implementation of the numerical algorithm.

\section*{Acknowledgment}
Part of this research was conducted during a visit of Zahra Khorsand to the University of California, Merced,
and the authors would like to express their gratitude for hospitality and support. 
HK and ZK also acknowledge support by the Research Council of Norway.
Dimitrios Mitsotakis was supported by the Marsden Fund administered by the Royal Society of New Zealand. 
Dimitrios Mitsotakis also thanks the participants of the ``Dispersive Hydrodynamics meeting at BIRS, 2015'', 
and especially Professors Gavin Esler, Sergey Gavrilyuk and Edward Johnson for fruitful discussions on the \acs{SGN} equations. 


\appendix

\section{The numerical method}\label{app:A}

In this Appendix we consider the \acf{IBVP} comprised of system (\ref{eq:DSerre1})--(\ref{eq:DSerre2}) subject to reflective boundary conditions. Rewriting the system in 
terms of $(h,\ u)$, and dropping the bar over the symbol of the horizontal velocity, yields the \acs{IBVP}
\begin{equation}\label{eq:serre2}
\begin{array}{l}
h_t + (hu)_x = 0\ , \\
hu_{t} + huu_{x} + g(h+b)_{x}+ \\
\left[ h^{2} \big( \Sfrac{1}{3} \PP + \Sfrac{1}{2} \QQ \big) \right]_x
+ b_x \big( \Sfrac{1}{2} \PP +  \QQ \big) =0\ ,  \\
u(a,t) =u(b,t)=0\ ,  \\
h(x,0) = h_0(x)\ ,  \\
u(x,0) = u_0(x)\ , 
\end{array} 
\end{equation}
where
$ \PP =  h \left[ u_{x}^{2} - u_{xt} - u u_{xx} \right]$, 
$ \QQ = b_x (u_t + uu_x) + b_{xx} u^2$, $x \in [a,\ b]\subset \R$ and $t \in [0,\ T]$. Considering a
spatial grid $x_i = a+i\ \Delta x$, for $i=0,1,\cdots, N$, where $\Delta
x$ is the spatial mesh-length, such that $\Delta x =
(b-a)/N$, $N\in \N$.  We define the space of cubic splines 
\begin{equation*}
  S = \left\{\phi\in C^2 [a,b]\Big| \phi|_{[x_i, x_{i+1}]} \in \P^{3},\ 0\leq i\leq N-1\right\}\ ,
\end{equation*}
where $\P^k$ is the space of polynomials of degree $k$. We also consider the space 
\begin{equation*}
S_0=S\cap \left\{\phi \in C([a,b])\Big| \phi(a)=\phi(b)=0  \right\}\ .
\end{equation*}
The basis functions of the space $S$ and $S_0$ consist of the usual B-splines described in \cite{Schultz1973}.

\begin{table*}[ht!]
\centering
\begin{tabular}{llclc}
\hline
$N$ & $E_0[H]$ & rate for $E_0[H]$ & $E_0[U]$ & rate for $E_0[U]$\\
\hline
 $300$ &  $0.1211\times 10^{-8}$  & --  &  $0.6127\times 10^{-11}$ & -- \\
 $320$ &  $0.9674\times 10^{-9}$  & $3.4793$ & $0.4733\times 10^{-11}$ & $3.9983$ \\
 $340$ &  $0.7836\times 10^{-9}$  & $3.4772$ & $0.3714\times 10^{-11}$ & $3.9999$ \\
 $360$ &  $0.6422\times 10^{-9}$  & $3.4797$ & $0.2955\times 10^{-11}$ & $3.9977$ \\
 $380$ &  $0.5322\times 10^{-9}$  & $3.4754$ & $0.2382\times 10^{-11}$ & $3.9885$ \\
 $400$ &  $0.4452\times 10^{-9}$  & $3.4793$ & $0.1939\times 10^{-11}$ & $4.0099$ \\
\hline 
\end{tabular}
\caption{Spatial errors and rates of convergence in the $L^2$ norm.}\label{tab:conergence1}
\end{table*}

The semi-discrete scheme is reduced in finding $\tilde{h}\in S$ and $\tilde u\in S_0$ such that
  \begin{equation}
  \begin{array}{l}
(\tilde{h}_t,\phi)+\left((\tilde{h}\tilde{u})_x,\phi \right)=0\ ,  \\
\mathcal{B}(\tilde{u}_t,\psi;\tilde{h})+\left(\tilde{h}\tilde{u}\tilde{u}_{x} + g(\tilde{h}+b)_{x},\psi \right)+\\
+
\left( \tilde{h}^{2} \big( \Sfrac{1}{3} \tilde{\PP} + \Sfrac{1}{2} \tilde{\QQ} \big), \psi_x \right)
+ \left( b_x ( \Sfrac{1}{2} \tilde{\PP} + \tilde{ \QQ} ),\psi \right)=0\ , 
  \end{array}\label{eq:semid}
  \end{equation}
for $\phi\in S$, and $\psi \in S_0$, and $\tilde{\PP} =  \tilde{h} \left[ \tilde{u}_{x}^{2} - \tilde{u} \tilde{u}_{xx} \right]$ and $ \tilde{\QQ} = b_x \tilde{u}\tilde{u}_x +
b_{xx} \tilde{u}^2$. $\mathcal{B}$ is defined as the bilinear form that for fixed
$\tilde{h}$ is given by 
\begin{multline}\label{eq:bilinear}
 \mathcal{B}(\psi,\chi;\tilde{h})= \left(\tilde{h}\left[1+\tilde{h}_xb_x+\frac{1}{2}\tilde{h}b_{xx}+b_x^2 \right]\psi,\chi \right)+\\ 
 +\third\left(\tilde{h}^3\psi_x,\chi_x \right)\ \mbox{for } \psi,\chi\in S_0\ .
\end{multline}
The system of equations (\ref{eq:semid}) is accompanied by the initial conditions 
\begin{equation}\label{eq:initconds}
  \tilde{h}(x,0) = \mathcal{P}\{h_0(x)\}\ ,\quad 
  \tilde{u}(x,0) = \mathcal{P}_0\{u_0(x) \}\ ,
\end{equation}
where $\mathcal{P}$ and $\mathcal{P}_0$ are the $L^2$-projections onto $S$ and $S_0$ respectively, satisfying
$(\mathcal{P}v,\phi) = (v,\phi)$ for all $\phi\in S$ and $(\mathcal{P}_0v,\psi) = (v,\psi)$ for all $\psi\in S_0$. Upon choosing
basis functions $\phi_j$ and $\psi_j$ for the spaces $S$ and $S_0$, (\ref{eq:semid}) is reduced to a system of
\acf{ODEs}. For the integration in time of this system we employ the
Dormand--Prince adaptive time-stepping methods,
\cite{Butcher2008, Hairer2009}. One may apply the same numerical method to solve the \acs{IBVP} with non-homogeneous Dirichlet boundary conditions. For example if $u(a,t)=u_a$ 
then the change of variables $u(x,t)=w(x,t)+u_0(x)$ reduces the non-homogeneous system to a homogeneous \acs{IBVP} system for the variable $w$. In all the numerical experiments 
we took $\Delta x=0.1$, while the tolerance for the relative error of the adaptive Runge--Kutta scheme was taken $5\cdot 10^{-14}$.
For the computations of the integrals the Gauss-Legendre quadrature rule with $8$ nodes was employed. 

The convergence properties of the standard Galerkin method for the \acs{SGN} system are very similar with those of the classical Boussinesq system studied in detail 
in \cite{AD2012,AD2013}. In order to compute the convergence rates in various norms we consider the nonhomogeneous \acs{SGN} system with flat bottom admitting the exact solution 
$h(x,t)= 1+e^{2t}(\cos(\pi x)+x+2)$ and $u(x,t)=e^{-tx}x\sin(\pi x)$ for $0\leq x\leq 1$, and for $t\in (0,T]$ with $T=1$. We compute the normalized errors
\begin{equation}\label{eq:errsnorm}
E_s[F]\doteq \frac{\|F(x,T;\Delta x)-F_{\mbox{exact}}(x,T)\|_s}{\|F_{\mbox{exact}}(x,T)\|_s}\ ,
\end{equation}
where $F=F(\cdot ;\Delta x)$ is the computed solution, i.e., either $H\approx h(x,T)$ or $U\approx u(x,T)$, $F_{\mbox{exact}}$ is the corresponding exact solution 
and $s=0,1,2,\infty$ correspond to the $L^2$, $H^1$, $H^2$ and $L^{\infty}$ norms, respectively.  The analogous rates of convergence are defined as
\begin{equation}
\mbox{rate for }E_s[F]\doteq \frac{\ln (E_s[F(\cdot;\Delta x_{k-1})]/E_s[F(\cdot;\Delta x_k)])}{\ln(\Delta x_{k-1}/\Delta x_k)}\ ,
\end{equation} 
where $\Delta x_k$ is the grid size listed in row $k$ in table \ref{tab:conergence1}. To ensure that the errors incurred by the temporal integration do not affect the rates of 
convergence we use $\Delta t\ll \Delta x$ while we take $\Delta x=1/N$.

Table \ref{tab:conergence1} presents the spatial convergence rates in the $L^2$ norm. We observe that the convergence is optimal for the $u$ variable but suboptimal 
for the $h$ variable. Specifically, it appears that $\|h-\tilde{h}\|\sim \Delta x^{3.5}$, while $\|u-\tilde{u}\|\sim\Delta x^{4}$. More precisely, as in the case of the classical Boussinesq system \cite{AD2012}, and because the rate of convergence in $h$ appears to be less that $3.5$ yields that the error should be of $O(\Delta x^{3.5}\sqrt{\ln (1/\Delta x)})$. 
Similar results obtained for the convergence in the $H^1$, $H^2$ and $L^\infty$ norms. Specifically it was observed numerically that $\|h-\tilde{h}\|_s\sim \Delta x^{3.5-s}$, 
$\|u-\tilde{u}\|_1\sim\Delta x^{4-s}$, for $s=0,1,2$ and $\|h-\tilde{h}\|_\infty\sim \Delta x^{3}$, while $\|u-\tilde{u}\|_\infty\sim\Delta x^{4}$ approximately.

\bibliographystyle{plain}

\end{document}